\documentclass[twocolumn,pra,aps,nofootinbib,superscriptaddress]{revtex4-1}
\usepackage[english]{babel}
\usepackage[utf8]{inputenc}
\usepackage{url,psfrag,graphicx,color}
\usepackage{amsmath,amssymb,amsthm}
\usepackage{bm}
\usepackage{hyperref}
\usepackage{physics}
\usepackage{epsfig}
\usepackage{mathptmx}
\usepackage[english]{babel}
\usepackage{amssymb}
\usepackage{amsfonts}
\usepackage{amsmath}
\usepackage{eucal}
\usepackage{graphicx,color,placeins}
\usepackage{epsfig}
\usepackage{tikz}
\usepackage{empheq}

\usepackage{cancel}
\usepackage{hyperref}
\usepackage{adjustbox}
\usepackage[normalem]{ulem}

\usepackage{makecell}

\def\<{\left<}
\def\>{\right>}
\def\ket|#1>{\left|#1\right>}
\def\bra<#1|{\left<#1\right|}
\def\Tr{\text{\rm Tr}}

\def\elem<#1|#2|#3>{\left<#1\right|#2\left|#3\right>}
\def\({\left(}
\def\){\right)}
\def\[{\left]}
\def\]{\right]}
\def\Z{{\mathbb Z}}
\def\C{{\mathbb C}}
\def\R{{\mathbb R}}
\def\N{{\mathbb N}}
\def\beq{\begin{equation}}
\def\eeq{\end{equation}}

\font\numbers=cmss12
\font\upright=cmu10 scaled\magstep1
\def\stroke{\vrule height8pt width0.4pt depth-0.1pt}
\def\topfleck{\vrule height8pt width0.5pt depth-5.9pt}
\def\botfleck{\vrule height2pt width0.5pt depth0.1pt}
\def\Zmath{\vcenter{\hbox{\numbers\rlap{\rlap{Z}\kern
0.8pt\topfleck}\kern 2.2pt
                   \rlap Z\kern 6pt\botfleck\kern 1pt}}}
\def\Qmath{\vcenter{\hbox{\upright\rlap{\rlap{Q}\kern
                   3.8pt\stroke}\phantom{Q}}}}
\def\Nmath{\vcenter{\hbox{\upright\rlap{I}\kern 1.7pt N}}}
\def\Cmath{\vcenter{\hbox{\upright\rlap{\rlap{C}\kern
                   3.8pt\stroke}\phantom{C}}}}
\def\Rmath{\vcenter{\hbox{\upright\rlap{I}\kern 1.7pt R}}}
\def\Hmath{\vcenter{\hbox{\upright\rlap{I}\kern 1.7pt H}}}
\def\Amath{\vcenter{\hbox{\upright\rlap{I}\kern 1.7pt A}}}
\def\Z{\ifmmode\Zmath\else$\Zmath$\fi}
\def\Q{\ifmmode\Qmath\else$\Qmath$\fi}
\def\N{\ifmmode\Nmath\else$\Nmath$\fi}
\def\C{\ifmmode\Cmath\else$\Cmath$\fi}
\def\R{\ifmmode\Rmath\else$\Rmath$\fi}

\def\ket|#1>{| #1 \rangle}
\def\bra<#1|{\langle #1 |}
\def\<{\langle}
\def\>{\rangle}
\def\{{\lbrace}
\def\}{\rbrace}
\def\({\left(}
\def\){\right)}
\def\[{\left[}
\def\]{\right]}

\def\be{\begin{equation}}
\def\ee{\end{equation}}
\def\bea{\begin{eqnarray}}
\def\eea{\end{eqnarray}}
\def\Tr{{\rm Tr}}

\def\ket|#1>{| #1 \rangle}
\def\bra<#1|{\langle #1 |}
\def\<{\langle}
\def\>{\rangle}
\def\{{\lbrace}
\def\}{\rbrace}
\def\({\left(}
\def\){\right)}

\def\beq{\begin{equation}}
\def\eeq{\end{equation}}
\def\barray{\begin{eqnarray}}
\def\earray{\end{eqnarray}}
\newcommand*\widefbox[1]{\fbox{\hspace{1em}#1\hspace{1em}}}
\begin{document}

\title{Simulating violation of causality  using a topological phase transition}
%\title{Entanglement spectrum and the string order parameter for the topological bilinear-biquadratic spin 1 Hamiltonian}  

\author{Sudipto Singha Roy}
\affiliation{Instituto de Física Teórica, UAM-CSIC, Universidad
  Aut{\'o}noma de Madrid, Cantoblanco, Madrid, Spain}
  \author{Anindita Bera}
\affiliation{Institute of Physics, Faculty of Physics, Astronomy and Informatics,
Nicolaus Copernicus University, Grudzi\c adzka 5/7, 87-100 Toru{\'n}, Poland}

\author{Germán Sierra}
\affiliation{Instituto de Física Teórica, UAM-CSIC, Universidad
  Aut{\'o}noma de Madrid, Cantoblanco, Madrid, Spain}

\date{\today}

\begin{abstract}
We consider a topological Hamiltonian and establish a correspondence between its eigenstates and the resource for a  causal order game introduced in Ref. \cite{brukner}, known as {\it process matrix}. We show that quantum correlations generated in the quantum many-body energy eigenstates of the model can mimic the statistics that can be obtained by exploiting different quantum measurements on the process matrix of the game. This provides an interpretation of the expectation values of the observables computed for the quantum many-body states in terms of the success probabilities of the game. As a result, we show that the ground state (GS) of the model can be related to the optimal strategy of the causal order game. Subsequently, we  observe that at the point of maximum violation of the classical bound in the causal order game, corresponding   quantum many-body model undergoes a second-order quantum phase transition (QPT).  The correspondence equally holds even when we generalize the game for a higher number of parties.
\end{abstract}
\maketitle

\section{Introduction}
Game-theoretic realization of quantum properties related to any physical system often provides a  better way of conceptualization of the underlying physical theory \cite{brukner, game_theory0, game_theory1,game_theory2,game_theory3,game_theory4,game_theory5,game_theory_entanglement1,
game_theory_entanglement2,game_theory_entanglement3,game_theory_nonlocality1,
game_theory_nonlocality2,game_theory6}.  For instance,  violation of 
Bell inequality  which is incompatible with the conjunction of  locality and realism can be formulated in the game-theoretic realm using
the Clauser–Horne–Shimony–Holt (CHSH) game \cite{CHSH}. 
 The key feature of any game theory consists of exploiting different strategies to optimize the cost function of the game.   In this regard, there have been studies where it is shown that for certain game theory set-up,  quantum strategies provide more advantages than their classical counterparts \cite{game_theory0,game_theory1,game_theory4}. This has led to further investigations for a deeper understanding of the role of entanglement \cite{acin_paper,game_theory_entanglement1,game_theory_entanglement2,game_theory_entanglement3} and nonlocality \cite{game_theory_nonlocality1,game_theory_nonlocality2} in any quantum game theory scheme. 
 
 A particularly interesting application of quantum many-player games could be to find its connection with quantum many-body systems. To begin with, we can think that different energy eigenstates of any quantum many-body Hamiltonian can be considered as strategies adopted by the quantum particles to attain a  configuration that satisfies the energy constraints.    Hence,  the total energy of the system resembles the cost function of any game theory scheme, and the GS of the model then corresponds to the optimal strategy adopted by the quantum particles to minimize the cost function of the game.  
 This motivates us to introduce a formalism that relates quantum many-body  Hamiltonians to an actual quantum game theory scheme in a more profound way.  In particular, we consider a topological Hamiltonian and show that the  energy eigenstates of the model can be related to the process matrix which is considered to be the main resource in the causal order game introduced in Ref. \cite{brukner}  by Oreshkov {\it et al}. We show that in this way, expectation values of certain non-commutative quantum observables computed for the quantum many-body eigenstates of the system can be interpreted as the success probabilities of the different strategies considered in the causal order game.  Moreover, we find that such a correspondence results in a classification of the eigenstates of the model based on the potentiality of violation of the classical bound by the process matrices to which the eigenstates can be related. Interestingly,  we find that the GS and the most excited state  of the model thus can be related to non-causally ordered process matrices that provide optimal success probability in the causal order game.  In addition to this, we identify that the maximum violation of classical bound in the causal order game corresponds to the point where in the thermodynamic limit, the quantum many-body model undergoes a second-order QPT. 

We organize the article as follows. In Sec. \ref{sec1} we introduce the quantum many-body Hamiltonian that we consider in our work. Thereafter,  in Sec. \ref{sec2} we summarize the key points of the causal order game. Sec. \ref{sec3} is devoted to introducing the formalism of our work and providing a correspondence between the quantum many-body system and quantum game theory scheme. In Sec. \ref{sec4}, we provide a generalization of the results to a higher number of parties. We discuss choices of relevant order parameters for identification of  QPT point in Sec. \ref{sec5}. We conclude and discuss future plans in Sec.\ref{sec6}.

\section{Model}
\label{sec1}
We start our discussion by introducing the quantum many-body Hamiltonian in one-dimension  that  will be the main focus in our work.  To have the initial set-up exactly similar to  the conventional causal order game for two parties, the minimum system size  we consider is $N=4$, for which the quantum many-body Hamiltonian reads as
\begin{eqnarray}
\mathcal{H}(\theta)=-2\cos\theta \sum_{i=1}^2\sigma_z^i \sigma_z^{i+2}-\sin\theta \sum_{i=1}^4 \sigma_z^i \sigma
_x^{i+1} \sigma_z^{i+2},
\label{eqn:hamiltonian}
\end{eqnarray}
where $\sigma_i^k$ are the Pauli matrices at site $k$ ($i \in x, y, z$) and we  consider periodic boundary conditions (PBC).  It is apparent that for this small system size the model can be diagonalized instantly.  However, one can note that even for any arbitrary $N$, the model can be exactly diagonalized by first applying certain non-local unitary transformation on pair of sites and then mapping it to a free-fermionic model. We discuss the methodology in detail in Sec. \ref{sec4} where we generalize the set-up for higher number of parties.
The Hamiltonian is translationally invariant and comprises certain symmetries. In particular, it commutes with the terms $\sigma_x^i \sigma_x^{i+2}$.  Now if we look at different parts of the Hamiltonian, then we can identify that $\sigma_z^i \sigma_z^{i+2}$ defines the Ising interaction between the non-nearest neighbor sites. Similarly, the second part,  $\sigma_z^i \sigma_x^{i+1} \sigma_z^{i+2}$ defines the  cluster Hamiltonian \cite{cluster_Ham1}  between nearest-neighbor sites. These two quantum Hamiltonians are characteristically very different from each other. In particular, the GS of cluster Hamiltonian with PBC  is non-trivial and it is also an example of symmetry-protected topological (SPT) state \cite{cluster_Ham2}. In the thermodynamic limit, the model exhibits a second-order QPT  at $\theta_c=\frac{\pi}{4}$ which we discuss in details in Sec.\ref{sec5}.

The main aim of our article is to provide a formalism to relate the  energy eigenstates of the above model to the resource of a  suitable quantum game theory scheme.  For that purpose,  we propose that the causal order game introduced in \cite{brukner}  (see also \cite{causal_order_exp,causal_order_more1,causal_order_more2,causal_order_more3,causal_order_more4}) can indeed be a potential candidate to establish such correspondence.  However, before going into the details of our formalism, we first briefly review the key points of the causal order game in the forthcoming section.

\section{Indefinite causal order revisited}
\label{sec2}
\begin{figure}[h]
\begin{center}
\includegraphics[width=7.cm,angle=-0]{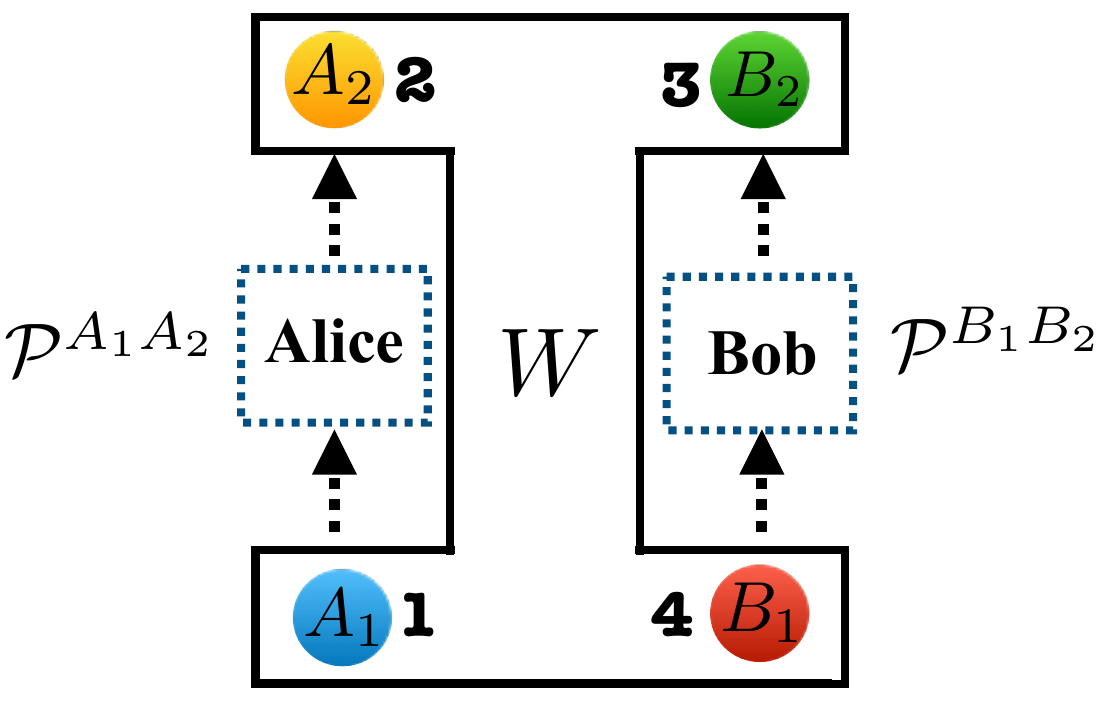}~~~
\end{center}
\caption{Schematic diagram of the arrangement of Alice's ($A_1,~A_2$) and Bob's  ($B_1,~B_2$)  quantum systems in the causal order game. The same arrangement  was also indexed by  1, 2, 3, 4 to explore its correspondence with a quantum many-body Hamiltonian  defined in  Eq. (\ref{eqn:hamiltonian}).}
\label{fig:schematic0}
\end{figure}

In this section, we revisit a causal order game between two observers Alice and Bob situated far from each other in their respective laboratories which are completely isolated from the external world.  Now at a given run of the game, each of them opens their laboratory once to  receive a particle ($A_1$  for Alice and $B_1$ for Bob) on which they can perform certain operations and later once to  send additional  systems ($A_2$ and $B_2$, respectively) out of their laboratories.   See schematic Fig. \ref{fig:schematic0} for the arrangements of Alice's and Bob's quantum systems. Now consider the following task to be performed by them. Once they receive the systems in their respective laboratories, each of the parties tosses a coin to obtain  random bits  `$a$' (for Alice)  and `$b$' (for Bob). The parties now have to guess each other's random bit  and  they will do that following the value of an additional random bit `$b'$' that Bob has to generate:  if $b'=0$, Bob will have to communicate the bit $b$ to Alice, whereas if $b'=1$, he will have to guess the bit $a$. Now let us denote Alice's and Bob's guess about each other's random bit ($a$ and $b$)   by $x$ and $y$, respectively. Hence, the game aims to maximize the success probability 
\begin{eqnarray}
P_{success}=\frac{1}{2} [P_{Alice}(x=b, b'=0)+P_{Bob}(y=a, b'=1)].
\end{eqnarray}
One can show that if all events obey causal order, no strategy can allow Alice and Bob to exceed the  classical bound $ P_{success} \leq P^{classical}_{success}=\frac{3}{4}$.

\begin{figure}[t]
\includegraphics[width=6.5cm,angle=-0]{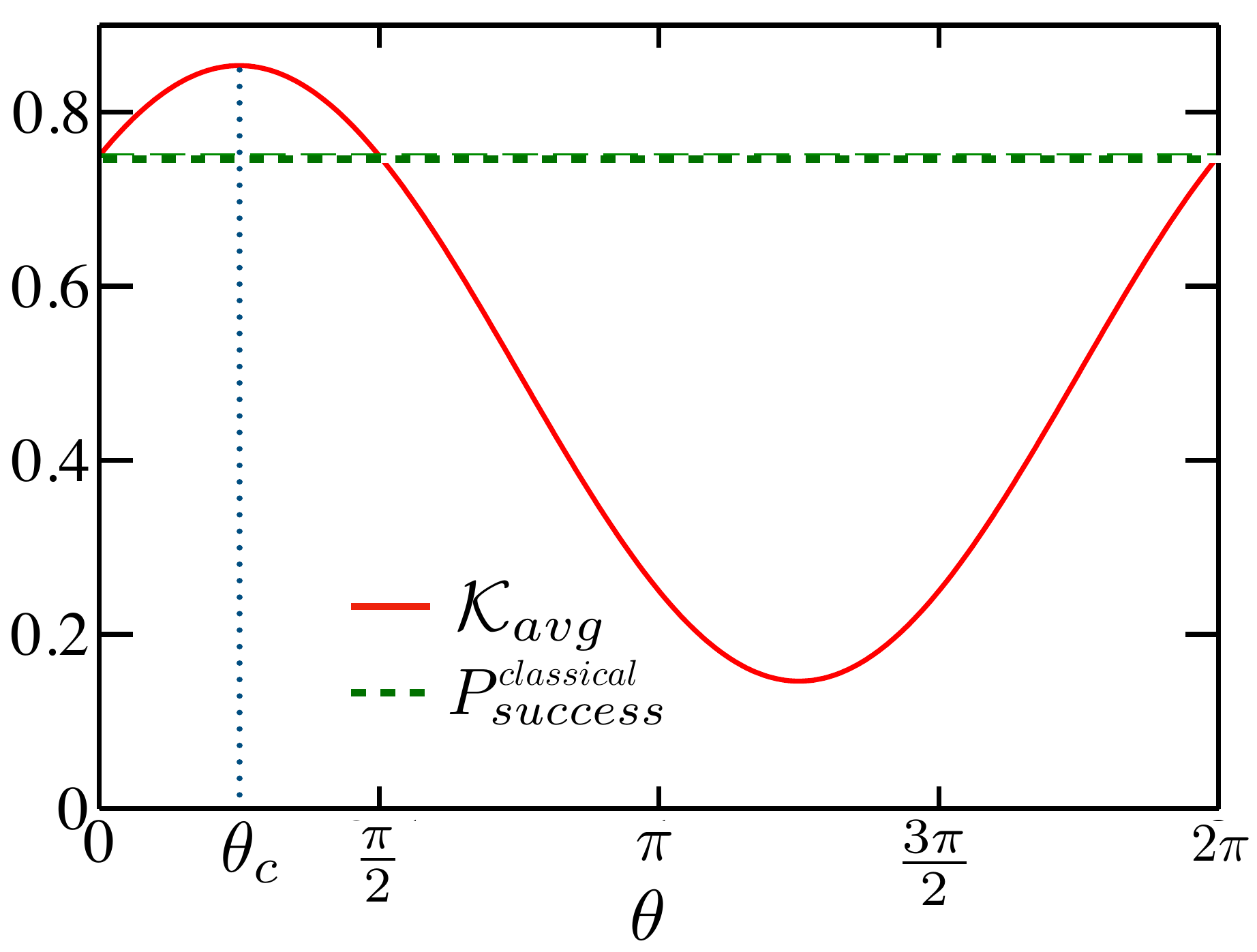}~~~
\caption{Plot of $\mathcal{K}_{avg}$ defined in Eq. (\ref{eqn:op_sum}) obtained for  the GS $|\Psi(\theta)\rangle_g$ of $\mathcal{H}(\theta)$ as a function of $\theta$ (solid red) which coincides with the total success probability of the causal order game as defined in Eq. (\ref{eq:success_prob_total}). The dashed green line corresponds to the  maximal classical bound, $P_{success}^{classical}=\frac{3}{4}$. We can see that $\mathcal{K}_{avg}$ (equivalently  $P_{success}$)  exceeds $P_{success}^{classical}$ for $0<\theta<\frac{\pi}{2}$. At $\theta_c=\frac{\pi}{4}$, the point of  maximal violation of causal bound, the observables $\Pi^0$ and $\Pi^1$ take same value $\frac{2+\sqrt{2}}{4}$. }
\label{fig:sum_observable}
\end{figure}

In Ref. \cite{brukner} it is shown that the bound can be violated if  we consider the following scenario, where the systems share a process matrix (see Ref \cite{process_mat_def}) given by  
\begin{eqnarray}
W_{opt}(\theta)=\frac{1}{4} \Big(\mathbb{I}^{A_1 A_2 B_1 B_2}+ \cos\theta\sigma^{A_2}_z \sigma^{B_1}_z+\sin\theta \sigma^{A_1}_z \sigma^{B_1}_x\sigma^{B_2}_z\Big), \nonumber\\
\label{eqn:process}
\end{eqnarray}
and apply certain measurement strategies.  Note that the case considered in  \cite{brukner} corresponds to $\theta =\frac{\pi}{4}$, where the violation is maximal.  Now the measurement strategies go as follows:  Alice always measures her input qubit in the $z$-basis and obtains the bit $x$ which is her guess about Bob's bit $b$. Thereafter, she encodes her random bit $a$ also in the $z$-basis. Therefore, the measurement operator in her part is given by 
\begin{eqnarray}
\mathcal{P}^{A_1A_2}_{b'}=\frac{1}{2}\Big(\mathbb{I}^{{A_1}}+(-1)^x\sigma_z^{A_1}\Big) \frac{1}{2}\Big(\mathbb{I}^{{A_2}}+(-1)^a\sigma_z^{A_2}\Big).
\label{eqn:Projector_Alice}
\end{eqnarray}
 On the other hand, Bob's measurement strategy has a dependence on the  bit  $b'$. If $b'=1$, he measures the input bit also in the $z$-basis obtaining $y$ which is his  guess about Alice’s bit $a$. In this case, how he encodes his bit is no longer important and we can denote the operator by $\rho^{B_2}$, with $\Tr(\rho^{B_2})=1$. However, when $b'=0$ he measures the input bit in the $x$-basis and  encodes the output as follows: if $y=0$, $b=0 \rightarrow |z_+^{B_2}\rangle$ and $b=1 \rightarrow |z_-^{B_2}\rangle$. Otherwise, if $y=1$, $b=1 \rightarrow |z_+^{B_2}\rangle$ and $b=0 \rightarrow |z_-^{B_2}\rangle$. Hence, the measurement operator in Bob's part reads as
\begin{eqnarray}
\mathcal{P}^{B_1B_2}_{b'}&=&b' \frac{1}{2}\Big(\mathbb{I}^{{B_1}}+(-1)^y\sigma_z^{B_1}\Big) \rho^{B_2}\nonumber\\&+&(1-b')\frac{1}{2}\Big(\mathbb{I}^{{B_1}}+(-1)^y \sigma_x^{B_1}\Big) \frac{1}{2}\Big(\mathbb{I}^{{B_2}}+(-1)^{y+b}\sigma_z^{B_2}\Big).\nonumber\\
\label{eqn:Projector_Bob}
\end{eqnarray}
In this way, when $b'=1$, a channel opens up between Alice's output and Bob's input and the success probability reads as 
{\small
\begin{eqnarray}
P_{Bob}(y=a,b'=1)&=&\sum_x \Tr\Big[ \frac{1}{2}\Big(\mathbb{I}^{{A_1}}+(-1)^x\sigma_z^{A_1}\Big) \frac{1}{2}\Big(\mathbb{I}^{{A_2}}+(-1)^a\sigma_z^{A_2}\Big) \nonumber\\
&&\frac{1}{2}\Big(\mathbb{I}^{{B_1}}+(-1)^y\sigma_z^{B_1}\Big) \rho^{B_2} W_{opt}\Big],\nonumber\\
&=&\frac{1+\cos\theta}{2}.
\label{prob:Bob}
\end{eqnarray}
}
Similarly, when $b'=0$, Bob  
 opens a channel with memory between his output and  Alice's input which helps  Alice to get the information of Bob's random bit b with probability
 {\small 
\begin{eqnarray}
P_{Alice}(x=b,b'=0)&=&\sum_y \Tr\Big[ \frac{1}{2}\Big(\mathbb{I}^{{A_1}}+(-1)^x\sigma_z^{A_1}\Big) \frac{1}{2}\Big(\mathbb{I}^{{A_2}}+(-1)^a\sigma_z^{A_2}\Big) \nonumber\\
&&\frac{1}{2}\Big(\mathbb{I}^{{B_1}}+(-1)^y\sigma_x^{B_1}\Big) \frac{1}{2}\Big(\mathbb{I}^{{B_2}}+(-1)^{y+b}\sigma_z^{B_2}\Big)\nonumber\\&&W_{opt}\Big]
=\frac{1+\sin\theta}{2}.
\label{prob:Alice}
\end{eqnarray}
}
Hence, the total success probability is given by
\begin{eqnarray}
P_{success}(\theta)&=&\frac{1}{2}\Big[P_{Alice}(x=b,b'=0)+P_{Bob}(y=a,b'=1)\Big], \nonumber\\
&=&\frac{1}{4}\Big[2+\cos\theta+\sin\theta\Big].
\label{eq:success_prob_total}
\end{eqnarray}
Therefore, the total success probability exceeds the classical bound $\frac{3}{4}$ for the region $0<\theta<\frac{\pi}{2}$
(see Fig. \ref{fig:sum_observable}). 

\section{Formalism}
\label{sec3}
 We devote this section to introduce the formalism necessary to establish the correspondence between  the quantum Hamiltonian expressed in Eq. (\ref{eqn:hamiltonian})   and    the causal order game introduced above. 
  Towards this aim,  we start with the  GS of the model defined in Eq. (\ref{eqn:hamiltonian}) which has the following  analytical form 
\begin{eqnarray}
|\Psi(\theta)\rangle_g&=& \cos^2\frac{\theta}{2}|\phi^+\rangle_{13}|\phi^+\rangle_{24}+\frac{\sin\theta}{2} \Big(|\phi^+\rangle_{12}|\psi^+\rangle_{34} \nonumber\\&+&|\psi^+\rangle_{12}|\phi^+\rangle_{34}\Big)-\sin^2\frac{\theta}{2} |\psi^+\rangle_{13}|\psi^+\rangle_{24},
\end{eqnarray}
where  $|\phi^{\pm}\rangle_{kl}=\frac{1}{\sqrt{2}}(|00\rangle_{kl}\pm|11\rangle_{kl})$ and $|\psi^{\pm}\rangle_{kl}=\frac{1}{\sqrt{2}}(|01\rangle_{kl}\pm|10\rangle_{kl})$ and  the sites have been indexed according to the schematic presented in Fig. \ref{fig:schematic0}. Therefore, when $\theta\rightarrow 0$, we can identify the GS  is the  Ising ferromagnet between next-nearest neighbor sites $|\Psi(0)\rangle_g=|\phi^+\rangle_{13}|\phi^+\rangle_{24}${\footnote{For $\theta\rightarrow0$ the GS of the Hamiltonian  is four-fold degenerate, comprises of the  states $|\phi^+\rangle_{13}|\phi^+\rangle_{24}$, $|\phi^+\rangle_{13}|\phi^-\rangle_{24}$, $|\phi^-\rangle_{13}|\phi^+\rangle_{24}$, and $|\phi^-\rangle_{13}|\phi^-\rangle_{24}$. However, in practice, there is always certain  external perturbation and  the system prefers one of them.}. Similarly, when $\theta \rightarrow \frac{\pi}{2}$, the GS becomes the cluster state $|\Psi(\frac{\pi}{2})\rangle_g=\frac{1}{4}\Pi_{i=1}^4 \Big(\mathbb{I}+\sigma_z^i \sigma_x^{i+1} \sigma_z^{i+2}\Big)|0000\rangle=\frac{1}{2}\Big(|0+0+\rangle+|0-1-\rangle+|1-0-\rangle+|1+1+\rangle\Big)$, with $|\pm\rangle=\frac{1}{\sqrt{2}}(|0\rangle \pm |1\rangle)$. 
We now argue that the expectation value of a set of observables\footnote{Note that due to translational symmetry, one could also choose $\Pi^0=\frac{1}{2}\big(\mathbb{I}^{123}+\sigma_z^1 \sigma_x^2 \sigma_z^3 \big)$, or $\Pi^0=\frac{1}{2}\big(\mathbb{I}^{234}+\sigma_z^2\sigma_x^3 \sigma_z^4 \big)$, or  $\Pi^0=\frac{1}{2}\big(\mathbb{I}^{124}+\sigma_x^1 \sigma_z^2 \sigma_z^4 \big)$, and $\Pi^1=\frac{1}{2}\big(\mathbb{I}^{13}+\sigma_z^1 \sigma_z^3\big)$.} computed for $|\Psi(\theta)\rangle_g$   can be related to the success probabilities of strategies employed on the process matrix given in Eq. (\ref{eqn:process})

\begin{widetext}

\begin{empheq}[box=\widefbox]{align}
&{_g}\langle \Psi(\theta)|\Pi^0|\Psi(\theta)\rangle_g =\frac{1+\sin\theta}{2} \leftrightarrow  P_{Alice}(x=b, b'=0)=\frac{1+\sin\theta}{2},\nonumber\\
&{_g}\langle \Psi(\theta)|\Pi^1|\Psi(\theta)\rangle_g =\frac{1+\cos\theta}{2}
  \leftrightarrow  P_{Bob}(y=a, b'=1)=\frac{1+\cos\theta}{2}, 
\label{eqn:observable_probability_correspondance}
\end{empheq}
\begin{eqnarray}
\text{where}~~ \Pi^0=\frac{1}{2}\big(\mathbb{I}^{134}+\sigma_z^1 \sigma_z^3 \sigma_x^4 \big), \Pi^1 &=&\frac{1}{2}\big(\mathbb{I}^{24}+\sigma_z^2 \sigma_z^4\big).
\label{eqn:operators}
\end{eqnarray}
 \end{widetext}

 This provides us with a scope to realize the  success probability of different parties in the causal order game as the expectation values  of  two non-commutative operators $\Pi^0$ and $\Pi^1$ computed for the GS. Hence, the average of the   correlators  resemble  the total success probability of the game (see Fig. \ref{fig:sum_observable}),
\begin{eqnarray}
\mathcal{K}_{avg}&=&\frac{{_g}\langle \Psi(\theta)|\Pi^0|\Psi(\theta)\rangle_g +{_g}\langle \Psi(\theta)|\Pi^1|\Psi(\theta)\rangle_g}{2}\nonumber\\&=&\frac{2+\cos\theta+\sin\theta}{4}.
\label{eqn:op_sum}
\end{eqnarray}
Therefore, we argue that the GS $|\Psi(\theta)\rangle_g$ corresponds to a causally non-separable process matrix when $\mathcal{K}_{avg}$ exceeds a minimum value,  $\frac{3}{4}$.  Moreover, one can note that the point of maximum violation of the classical bound, $\theta_c=\frac{\pi}{4}$, corresponds to the QPT point of the model.  This will be clear when we extend the model for a large system size. We discuss this in detail in Sec. \ref{sec5}.

A closer look at the three-body reduced density matrix derived from   $|\Psi(\theta)\rangle_{g}$ reveals the reason for the operators $\Pi^i$'s to take the above values. 
{\small
\begin{eqnarray}
\rho^{134}_g&=&\frac{1}{8}\Big[\mathbb{I}^{134}+ \cos \theta \sin\theta\sum_{i=1, 3, 4} \sigma_x^i\nonumber+ \cos \theta (\sigma_z^1 \sigma_z^3-\sigma_y^1 \sigma_y^3)\nonumber\\&+&\sigma_x^1 \sigma_x^3+ \cos\theta \sin\theta \sigma_x^1 \sigma_x^3 \sigma_x^4+\sin\theta (\sigma_z^1 \sigma_z^3 \sigma_x^4-\sigma_y^1 \sigma_y^3 \sigma_x^4)\Big]. \nonumber\\
\label{eqn:red_den}
\end{eqnarray}}
From the above expression, we can easily see that the $zz$-correlation between sites 1 and 3 (equivalently between site 2 and 4) and the three-body correlation $zzx$ between sites 1 3 4 (equivalently correlation $zxz$  between sites (1 2 3), (2 1 4) and (2 3 4)) take the values $\cos\theta$ and $\sin\theta$, respectively,  and makes  $\rho^{134}_g$ structurally equivalent to $W_{opt}$, as expressed in Eq. (\ref{eqn:process}).  However, $\rho^{134}_g$ consist of more terms than $W_{opt}$. This naturally leads to the following question: What would be the correspondence for the other  observables computed for the GS and other excited states of the model?

We now establish the relation between other excited states of the Hamiltonian $\mathcal{H}(\theta)$  together with the quantum observables computed for them to that of the strategies applied in the causal order game  using the formalism we propose below.  
 
 \begin{figure}[t]
\includegraphics[width=8.5cm,angle=-0]{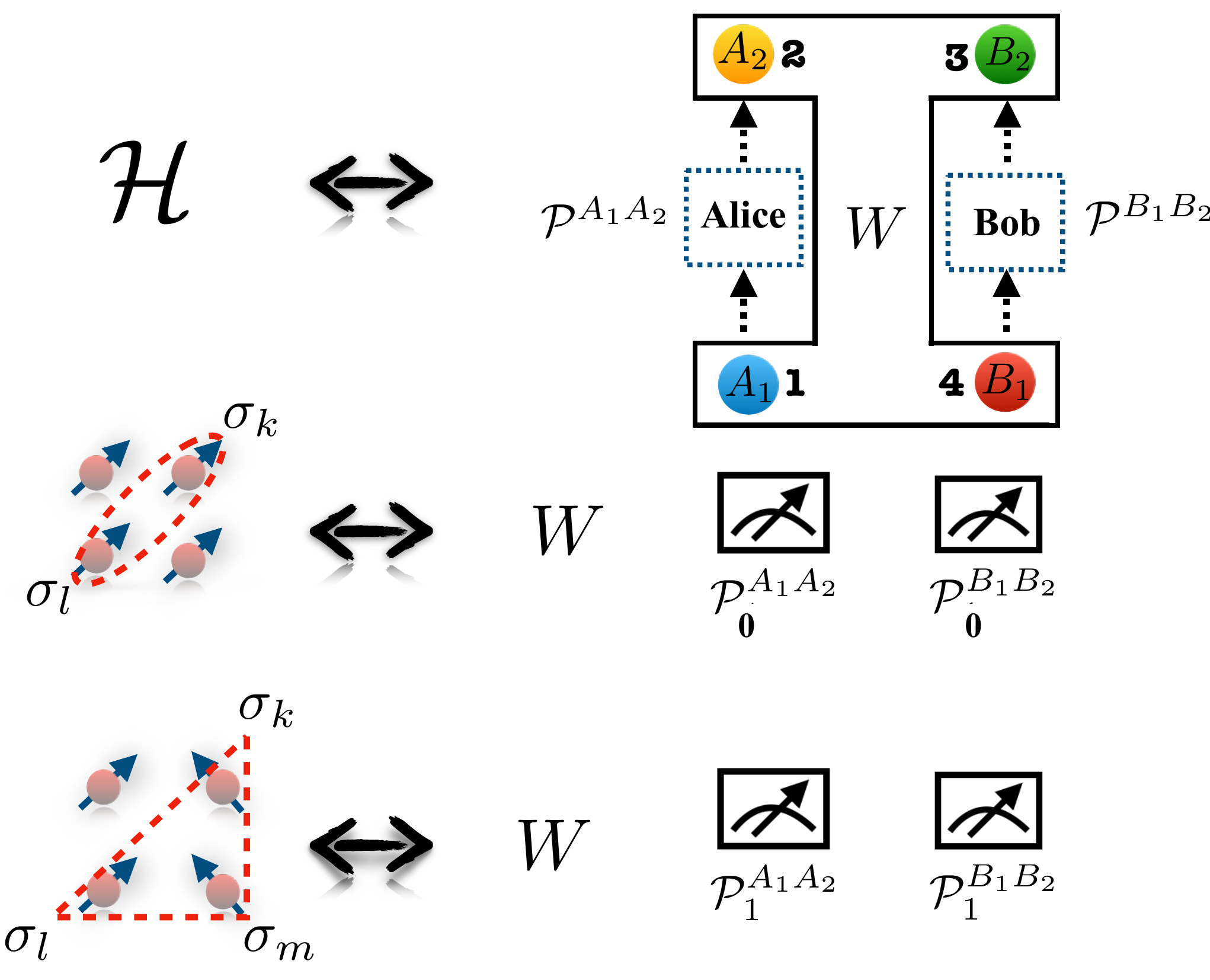}~~~
\caption{Schematic diagram of the analogy between quantum strategies exploited  in causal order game that includes process matrix and quantum measurements   and  the  strategies adopted by the quantum particles that consist of energy eigenstates and choice of certain physical observables.}
\label{fig:schematic1}
\end{figure}  
 
  {i) \it Quantum many-body Hamiltonian and quantum game theory scheme}: We first provide a realization of any quantum many-body Hamiltonian in terms of a quantum game theory scheme. In particular, we argue that the total energy of the system can be related to the pay-offs of any game theory.   Therefore, the eigenstates of the model can be considered as different strategies adopted by the quantum particles to yield a particular value of energy. In this way, we can think that the GS corresponds to the optimal strategy applied by the quantum particles to minimize the total energy of the system or the pay-off function of the game.   This lays out the initial set-up we need to provide a systematic comparison between a quantum many-body  Hamiltonian and an actual game theory scheme.  However, to relate the quantum many-body Hamiltonian $\mathcal{H}$ to the causal order game in a profound way, we need some additional aspects which we state in the next two axioms. 
  
 {ii) \it Eigenstates of $\mathcal{H}(\theta)$ and process matrix of causal order game:} 
In the case of the causal order game, as described above, all the participants agree on performing certain  quantum measurements on the process matrix to optimize the success probability, which we call the strategies of the game. When those strategies are applied, quantum channels between different parts of the system may open up which essentially assists to communicate the information about the random bit to be guessed.  In the quantum many-body systems, an analogy of this can be given by generalizing the notion of strategy introduced in the axiom i) as follows. When the quantum particles adopt a particular configuration satisfying the energy constraint, correlations may generate in different parts of the system. Hence,   the notion of strategy in quantum many-body systems comprises of two constituents,  the energy eigenstates and the choices of quantum operators quantifying different correlations in their subparts.   
 A quantitative way of conceptualization of this correspondence is stated in the axiom below.

  {iii) \it Quantum observables and success probability:}   We propose that the expectation values of the relevant physical operators computed  for the  quantum many-body eigenstates of the model can be related to the success probability of different strategies of the causal order game. 
 
 We are now ready with the necessary tools to relate the GS and quantum observables computed for it with that of different strategies of the causal order game. We summarize the correspondence in Table \ref{Table1} and a schematic representation of the same is  presented in Fig. \ref{fig:schematic1}.  For instance, in the first row of Table \ref{Table1}, we now establish correspondence between the expectation values of correlators $\sigma_y^1\sigma_y^3$, $\sigma_y^1\sigma_y^3 \sigma_x^4$  and the probabilities obtained from the causal order game. For  that purpose, similarly to  Eq. (\ref{eqn:observable_probability_correspondance}), we compute the expectation values of  two sets of observables $\Pi^0=\frac{1}{2}(\mathbb{I}^{134}+\sigma_y^1\sigma_y^3 \sigma_x^4)$ and $\Pi^1=\frac{1}{2}(\mathbb{I}^{13}+\sigma_y^{1}\sigma_y^{3})$ and compare it with the probabilities that can be obtained by applying the set of operators $\mathcal{P}^{A_1A_2}_{0(1)}$ and $\mathcal{P}^{B_1B_2}_{0(1)}$ defined in the table   on a process matrix given by {\small$W=\frac{1}{4}\Big(\mathbb{I}^{A_1A_2B_1B_2}-\cos\theta \sigma_y^{A_2} \sigma_y^{B_1}-\sin\theta \sigma_y^{A_1} \sigma_x^{B_1} \sigma_y^{B_2}\Big)$}. One can realize that the outcomes coincide with that obtained in Eq. (\ref{eqn:observable_probability_correspondance}), for $W_{opt}(\theta+\pi)$.

\begin{widetext}
\begin{center}
\begin{table}[h]
\begin{tabular}{ |l|l|l|}
\hline
 \multicolumn{1}{|c|}{No.} & \multicolumn{1}{c|}{Strategy} & \multicolumn{1}{c|}{Outcome}  \\\hline
% 1&$\begin{array}  {lcl}
1&  $\begin{array}  {lcl}
 \begin{cases}
|\Psi\rangle_g, \\  \Pi^0=\frac{1}{2}(\mathbb{I}^{134}+\sigma_y^1 \sigma_y^3 \sigma_x^4), \Pi^1=\frac{1}{2}(\mathbb{I}^{13}+\sigma_y^1 \sigma_y^3), 
\end{cases}
\\
\hspace{0.72cm}\updownarrow \\
 \begin{cases}
W=\frac{1}{4} \Big(\mathbb{I}^{A_1 A_2 B_1 B_2}- \cos\theta\sigma^{A_2}_y \sigma^{B_1}_y-\sin\theta \sigma^{A_1}_y\sigma^{B_1}_x\sigma^{B_2}_y\Big),\\

\mathcal{P}^{A_1A_2}_0 = \frac{1}{2}\Big(\mathbb{I}^{A_1}+(-1)^x\sigma_y^{A_1}\Big) \frac{1}{2}\Big(\mathbb{I}^{A_2}+(-1)^a\sigma_y^{A_2}\Big), \\\mathcal{P}^{B_1B_2}_0=\frac{1}{2}\Big(\mathbb{I}^{B_1}+(-1)^y \sigma_x^{B_1}\Big) \frac{1}{2}\Big(\mathbb{I}^{B_2}+(-1)^{y+b}\sigma_y^{B_2}\Big),\\

\mathcal{P}^{A_1A_2}_1= \frac{1}{2}\Big(\mathbb{I}^{A_1}+(-1)^x\sigma_y^{A_1}\Big) \frac{1}{2}\Big(\mathbb{I}^{A_2}+(-1)^a\sigma_y^{A_2}\Big), \\\mathcal{P}^{B_1B_2}_1=\frac{1}{2}\Big(\mathbb{I}^{B_1}+(-1)^y \sigma_y^{B_1}\Big) \rho^{B_2}. \\

\end{cases}
\end{array}$& $\begin{array}  {lcl}
{_g} \langle \Psi|\Pi^0\Psi\rangle_g &=&\frac{1-\sin(\theta)}{2}\\
 &\updownarrow& \\   P_{Alice}(x=b, b'=0)&=&\sum_y \Big[\text{Tr}\Big(\mathcal{P}_0^{A_1A_2} \mathcal{P}_0^{B_1B_2}W\Big)\Big]_{x=b}=\frac{1-\sin\theta}{2},\\\\
 {_g} \langle \Psi|\Pi^1|\Psi\rangle_g&=&\frac{1-\cos\theta}{2}\\
 &\updownarrow& \\   P_{Bob}(y=a, b'=1)&=& \sum_x \Big[\text{Tr}\Big(\mathcal{P}_1^{A_1A_2} \mathcal{P}_1^{B_1B_2}W\Big)\Big]_{y=a}=\frac{1-\cos\theta}{2} 
 \end{array} 
 $
\\
\hline
2&$\begin{array}  {lcl}
 \begin{cases}
|\Psi\rangle_g , \\ \Pi^0=\frac{1}{2}\mathbb{I}^{1234},  \Pi^1=\frac{1}{2}(\mathbb{I}^{13}+\sigma_x^1 \sigma_x^3),
\end{cases}
\\
\hspace{0.72cm}\updownarrow \\
 \begin{cases}
W=\frac{1}{4}\Big(\mathbb{I}^{A_2B_1}+\sigma_x^{A_2} \sigma_x^{B_1}\Big),\\
\mathcal{P}^{A_1A_2}_1 = \frac{1}{2}\Big(\mathbb{I}^{A_1}+(-1)^x\sigma_z^{A_1}\Big) \frac{1}{2}\Big(\mathbb{I}^{A_2}+(-1)^a\sigma_x^{A_2}\Big), \\\mathcal{P}^{B_1B_2}_1=\frac{1}{2}\Big(\mathbb{I}^{B_1}+(-1)^y\sigma_x^{B_1}\Big) \rho_{B_2}.
\end{cases}
\end{array}$
&  $\begin{array}  {lcl}
{_g} \langle \Psi|\Pi^0|\Psi\rangle_g&=&\frac{1}{2}
\\  &\updownarrow& \\   P_{Alice}(x=b, b'=0)&=&\frac{1}{2},\\\\
{_g} \langle \Psi|\Pi^1|\Psi\rangle_g&=&1
\\  &\updownarrow& \\   P_{Bob}(y=a, b'=1)&=&\sum_x \Big[\text{Tr}\Big(\mathcal{P}_1^{A_1A_2} \mathcal{P}_1^{B_1B_2}W\Big)\Big]_{y=a}=1
\end{array} 
$
\\ \hline

3&  $\begin{array}  {lcl}
 \begin{cases}
|\Psi\rangle_g, \\ \Pi^0=\frac{1}{2}(\mathbb{I}^{134}+\sigma_x^1 \sigma_x^3 \sigma_x^4),  \Pi^1=\frac{1}{2}(\mathbb{I}^{1234}), 
\end{cases}
\\
\hspace{0.72cm}\updownarrow \\
 \begin{cases}
W=\frac{1}{4} \Big(\mathbb{I}^{A_1 A_2 B_1 B_2}+\sin\theta \cos\theta \sigma^{A_1}_x\sigma^{B_1}_x\sigma^{B_2}_x\Big),\\
\mathcal{P}^{A_1A_2}_0=\frac{1}{2}\Big(\mathbb{I}^{A_1}+(-1)^x\sigma_x^{A_1}\Big) \frac{1}{2}\Big(\mathbb{I}^{A_2}+(-1)^a\sigma_x^{A_2}\Big), \\\mathcal{P}^{B_1B_2}_0=\frac{1}{2}\Big(\mathbb{I}^{B_1}+(-1)^y \sigma_x^{B_1}\Big) \frac{1}{2}\Big(\mathbb{I}^{B_2}+(-1)^{y+b}\sigma_x^{B_2}\Big). 
\end{cases}
\end{array}$&$\begin{array}  {lcl}
{_g} \langle \Psi|\Pi^0|\Psi\rangle_g &=&\frac{1+\sin\theta \cos\theta}{2}\\
 &\updownarrow& \\    P_{Alice}(x=b, b'=0)&=&\sum_y  \Big[\text{Tr}\Big(\mathcal{P}_0^{A_1A_2} \mathcal{P}_0^{B_1B_2}W\Big)\Big]_{x=b}=\frac{1+\sin\theta\cos\theta}{2},\\\\
 
{_g} \langle \Psi|\Pi^1|\Psi\rangle_g &=&\frac{1}{2}\\
 &\updownarrow& \\    P_{Bob}(y=a, b'=1)&=&\frac{1}{2} 
 \end{array}$\\
 \hline
\end{tabular}
\caption{Correspondence between strategies applied in causal order game  to that of the strategies employed in quantum-many body systems. Here, we consider the GS of the model.}
\label{Table1}
\end{table}
\end{center}
\end{widetext}

For the remaining correlators $\sigma_x^1\sigma_x^3$ and $\sigma_x^1\sigma_x^3 \sigma_x^4$, we need to modify the causal order game to some extent. For example, for $\sigma_x^1\sigma_x^3$, when $b'=1$ Alice assists Bob to guess her qubit like as the previous cases and the strategy now  involves the operators $\mathcal{P}^{A_1A_2}_1$ and $\mathcal{P}^{B_1B_2}_1$ defined in the table, and the process matrix given by  {\small $W=\frac{1}{4}\Big(\mathbb{I}^{A_2B_1}+\sigma_x^{A_2} \sigma_x^{B_1}\Big)$}. This results in   $P_{Bob}(y=a, b'=1)=1$,  which coincides with the expectation value of the operator $\Pi^1=\frac{1}{2}(\mathbb{I}^{13}+\sigma_x^1 \sigma_x^3)$.  However,  when $b'=0$, Bob  becomes  biased and does not want   to communicate his random bit $b$ to Alice. Hence,  Alice can guess  Bob's random bit with a probability $P_{Alice}(x=b, b'=0)=\frac{1}{2}$, which coincides with the expectation value of the opertaor $\Pi^0=\frac{1}{2}\mathbb{I}^{1234}$. Therefore, in this case, the game consists of  effectively only one  quantum strategy ($b'=1$) and the total success probability thus turns out to be {\small $P_{success}=\frac{1}{2}\Big[P_{Alice}(x=b, b'=0)+P_{Bob}(y=a, b'=1)\Big]=\frac{3}{4}$}.  Similarly, for the three-body correlator  $\sigma_x^1\sigma_x^3 \sigma_x^4$ we consider the opposite scenario, i.e., in this case, Bob always assists Alice to guess his random qubit $b$ ($b'=0$) and the strategy consists of measurement operators $\mathcal{P}^{A_1A_2}_0$ and $\mathcal{P}^{B_1B_2}_0$ defined in the table and the process matrix given by  {\small $W=\frac{1}{4} \Big(\mathbb{I}^{A_1 A_2 B_1 B_2}+\sin\theta \cos\theta \sigma^{A_1}_x\sigma^{B_1}_x\sigma^{B_2}_x\Big)$}, which yields $P_{Alice}(x=a, b'=0)=\frac{1+\sin\theta\cos\theta}{2}$ and coincides with the expectation value of the operator $\Pi^0=\frac{1}{2}(\mathbb{I}^{134}+\sigma_x^1 \sigma_x^3 \sigma_x^4)$. However, when $b'=1$, Alice  becomes biased and does not help Bob to guess her random bit $a$. Therefore,  Bob  can only guess about Alice's bit randomly, with a probability $P_{Bob}(y=a, b'=1)=\frac{1}{2}$  that matches with the expectation value of $\Pi^1=\frac{1}{2}(\mathbb{I}^{1234})$. Hence, the game again involves effectively only one quantum strategy ($b'=0$) and the total success probability becomes  {\small $P_{success}=\frac{1}{2}\Big[P_{Alice}(x=b, b'=0)+P_{Bob}(y=a, b'=1)\Big]=\frac{2+\sin\theta \cos\theta}{4}$}, which  remains smaller than the classical bound $\frac{3}{4}$ for all values of $\theta$. 
\vspace{-0.0145cm}

In addition to this, one can now realize   that  the correspondence shown in Eq. (\ref{eqn:observable_probability_correspondance}) also holds for  the most excited state of  $\mathcal{H}$, for which   we get $\rho_{ex}^{134}(\theta)=\rho_g^{134}(\theta+\pi)$ as a manifestation of  $\mathcal{H(\pi+\theta)}=-\mathcal{H}(\theta)$.  Therefore, for the same choices of $\Pi^0$ and $\Pi^1$ as defined in Eq. (\ref{eqn:operators}), the most excited state of the model ($|\Psi\rangle_{ex}$) can also be related to the process matrix $W_{opt}(\theta+\pi)$ that violates the classical bound in the causal order game (for $\pi< \theta<\frac{3\pi}{2}$).  Along with this, one can establish a correspondence between all the other observables of $|\Psi\rangle_{ex}$ and the strategies in the causal order game again by using Table \ref{Table1} and doing the transformation  $\theta \rightarrow \theta+\pi$. We carry out the same exercise for all other excited states of the model $\mathcal{H}$. However, we report that such a correspondence fails to relate any of them to a process matrix that can violate the classical bound for any value of the parameter $\theta$. Hence, a classification between the eigenstates of the model emerges, where the GS and most excited state are considered to be in the same set as they can be related to a non-causal ordered process matrix. Whereas, all other remaining excited states of the model belong to the complementary set. \\

\begin{figure}[t]
\begin{center}
\includegraphics[width=5.cm,angle=-0]{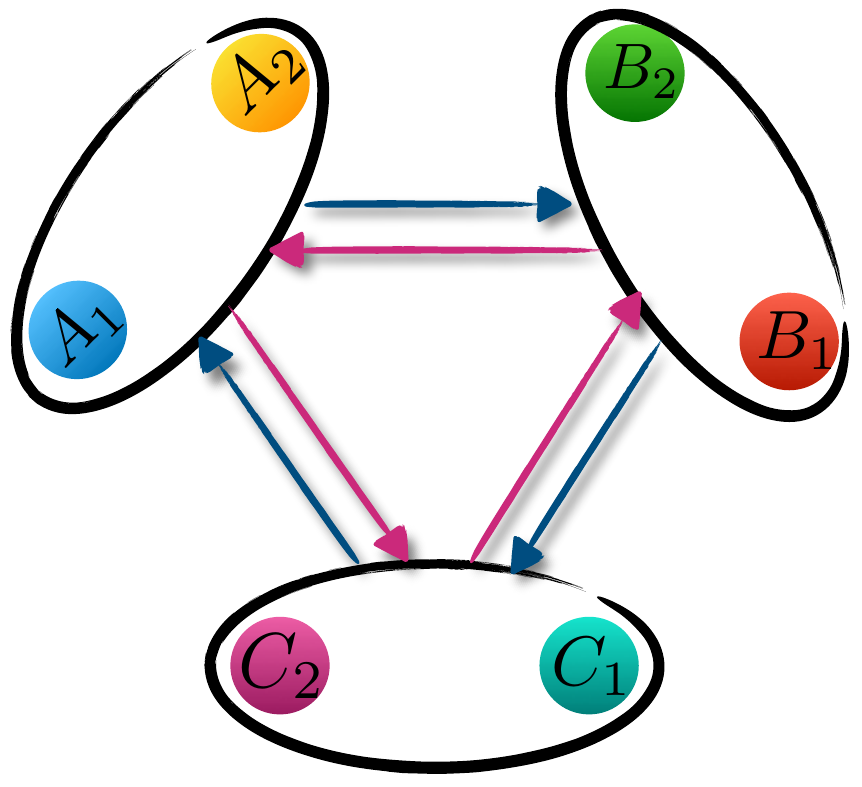}~~~
\end{center}
\caption{Schematic diagram of the arrangement of Alice's ($A_1, ~A_2$), Bob's  ($B_1, ~B_2$), and  Charlie's  ($C_1, ~C_2$)  quantum systems in the  three-party causal order game. Each  arrow connects a pair of parties such that the start point of the arrow indicates the party that will generate the random bit ($a$ or $b$ or c). Whereas, the end point of the arrow indicates the party that will produce a guess bit ($x$ or $y$ or $z$) of that random bit.}
\label{fig:schematic2}
\end{figure} 

\section{Generalization to higher parties}
\label{sec4}
In this section, we provide a methodology to generalize the causal order game presented above for a higher number of parties keeping the set-up compatible with the quantum many-body Hamiltonian that we consider in our work. In particular,  we generalize the above game for three parties ($N=6$) and derive the corresponding classical and quantum bound of the success probabilities. This methodology could be useful to extend the game for any number of parties. Consider the game now consists of three parties Alice, Bob, and Charlie with their input and output systems denoted by $(A_1, A_2), (B_1, B_2)$, and $(C_1, C_2)$, respectively. Additionally, we consider Alice is  the left neighbor of Bob and the right neighbor of Charlie. Similarly Bob is the  left neighbor of Charlie and the right neighbor of Alice. Finally, Charlie is the left neighbor of Alice and the right neighbor of Bob.  Moreover, similar to the two-player game, the random bit generated by the parties can be denoted as $a$ (for Alice), $b$ (for Bob), and $c$ (for Charlie).  Each of the parties now has to guess the random bit of the other two parties (see schematic presented in Fig. \ref{fig:schematic2}) and they will do that following the value of an additional random bit `$b'$' that will be generated  by an external agent Crupier:   if $b'=0$, the parties has to guess the bit of their left neighbor. Whereas, if $b'=1$, each of the party will guess the bit of its right neighbor.  Now if we denote the guess bits produced by all the parties as $x$ (for Alice), $y$ (for Bob), and $z$ (for Charlie), the task will be to maximize the probability function given by

\begin{eqnarray}
P_{success}&=&\frac{1}{2}\Big[P^{Left}_{success}+ P^{Right}_{success}\Big], 
\label{eqn:prob_new}
\end{eqnarray}
with 
\begin{eqnarray}
P_{success}^{Left}(b'=0)&=& \frac{1}{3}\Big[P_{Alice}(x=c)+P_{Bob}(y=a)\nonumber\\&+&P_{Charlie}(z=b)\Big],\\
 P_{success}^{Right}(b'=1)&=&\frac{1}{3}\Big[P_{Alice}(x=b)+ P_{Bob}(y=c)\nonumber\\&+&P_{Charlie}(z=a)\Big].
\label{eqn:prob_new}
\end{eqnarray}
Now let us consider there exist a global causal order such as   $A \preceq B \preceq C$. In that case, one can show that $P_{success}$ is bounded above by $3/4$. This can be proved as follows. As Alice is in the causal past of both Bob and Charlie, Alice can guess their bits with maximum probability $P_{Alice}(x=b)=P_{Alice}(x=c)=\frac{1}{2}$. Whereas, Bob is in the causal past of Charlie and in the future of Alice. Hence, with the help of Alice, Bob can guess her bit perfectly but can only guess the bit of Charlie, randomly. Hence, we get    $P_{Bob}(y=a)=1$, and $P_{Bob}(y=c)=\frac{1}{2}$. However, Charlie remains in the causal future of both Alice and Bob. Hence, he can guess their bits perfectly  $P_{Charlie}(z=a)=P_{Charlie}(z=b)=1$.  Therefore, we finally get 
\begin{eqnarray}
P^{Right}_{success}&=&\frac{1}{3}\Big[\frac{1}{2}+\frac{1}{2}+1]=\frac{2}{3}, 
P^{Left}_{success}=\frac{1}{3}\Big[\frac{1}{2}+1+1\Big]=\frac{5}{6}, \nonumber\\
\end{eqnarray}
which yields $P_{success}=\frac{3}{4}$.

To formulate a quantum version, we take the following process matrix
\begin{eqnarray}
&&W(\theta)=\frac{1}{8}\Big[\mathbb{I}+\frac{f_0(\theta, ~N=6)}{3}(\sigma_z^{A_2}\sigma_z^{B_1}+\sigma_z^{B_2}\sigma_z^{C_1}+\sigma_z^{A_1}\sigma_z^{C_2})\nonumber\\&+&\frac{f_1(\theta, ~N=6)}{3}(\sigma_x^{A_1} \sigma_x^{B_1}\sigma_z^{B_2}+\sigma_x^{B_1}\sigma_x^{C_1}\sigma_z^{C_2}+\sigma_x^{A_1}\sigma_z^{A_2}\sigma_x^{C_1})\Big],\nonumber\\
\label{eqn:process_three_party}
\end{eqnarray}
where the meaning of the functions $f_0(\theta, ~N=6)$ and $f_1(\theta,~N=6)$ will be clear when we introduce corresponding quantum many-body Hamiltonian. }
Let us now consider  the measurement operators for input and  output systems of each of the parties that can be expressed as
\begin{eqnarray}
{\mathcal P}^{\chi_1\chi_2}_{b'}(\alpha, \beta; \zeta)&=&\frac{1}{2} (\mathbb{I}^{\chi_1}+(-1)^{\alpha}\sigma_{\zeta}^{\chi_1}) \frac{1}{2} (\mathbb{I}^{\chi_2}+(-1)^{\beta}\sigma_{z}^{\chi_2}),\nonumber\\
\end{eqnarray}
where $\chi \in A, B, C$, and $\zeta \in x, y, z$. 
The quantum measurement operators for each parties for a certain value of $b'$  and the probabilities  are summarized in Table \ref{Table2}. The total projector ${\mathcal P}^{A_1A_2B_1B_2C_1C_2}_{b'}(\alpha, \beta; \zeta)$ is the product of the individual $\mathcal P$'s given in the table. 
\begin{widetext}
\begin{center}
\begin{table}[h]
\begin{tabular}{ |l|l|l|l|l|}
\hline
 \multicolumn{1}{|c|}{$b'$} & \multicolumn{1}{c|}{$A$} & \multicolumn{1}{c|}{$B$}  & \multicolumn{1}{c|}{$C$}  & \multicolumn{1}{c|}{$Probability$}  \\\hline
% 1&$\begin{array}  {lcl}
0&  $\begin{array}  {lcl}
 {\mathcal P}^{A_1A_2}_{b'=0}(x, 0; z)
\end{array}$
& 
$\begin{array}  {lcl}
 {\mathcal P}^{B_1B_2}_{b'=0}(0, 0; z)
 \end{array} $
 &
 $\begin{array}  {lcl}
 {\mathcal P}^{C_1 C_2}_{b'=0}(0, c; z)
 \end{array} $
 &  $\begin{array}  {lcl}
P_{Alice}(x=c, b'=0)= \frac{1+f_0((\theta,~N=6)}{2}
 \end{array} $\\\hline
0&  $\begin{array}  {lcl}
 {\mathcal P}^{A_1A_2}_{b'=0}(0,a; z)
\end{array}$
& 
$\begin{array}  {lcl}
 {\mathcal P}^{B_1B_2}_{b'=0}(y, 0; z)
 \end{array} $
 &
 $\begin{array}  {lcl}
 {\mathcal P}^{C_1C_2}_{b'=0}(0, 0;  z)
 \end{array} $
 &  $\begin{array}  {lcl}
P_{Bob}(y=a, b'=0)= \frac{1+f_0(\theta, ~N=6)}{2}
 \end{array} $\\\hline
  0&  $\begin{array}  {lcl}
 {\mathcal P}^{A_1A_2}_{b'=0}(0, 0; z)
\end{array}$
& 
$\begin{array}  {lcl}
 {\mathcal P}^{B_1 B_2}_{b'=0}(0, b; z)
 \end{array} $
 &
 $\begin{array}  {lcl}
 {\mathcal P}^{C_1 C_2}_{b'=0}(z, 0; z)
 \end{array} $
 &  $\begin{array}  {lcl}
P_{Charlie}(z=b, b'=0)= \frac{1+f_0(\theta,~N=6)}{2}
 \end{array} $\\\hline
1&  $\begin{array}  {lcl}
 {\mathcal P}^{A_1A_2}_{b'=1}(x, x; x)
\end{array}$
& 
$\begin{array}  {lcl}
 {\mathcal P}^{B_1B_2}_{b'=1}(0, b; x)
 \end{array} $
 &
 $\begin{array}  {lcl}
 {\mathcal P}^{C_1 C_2}_{b'=1}( 0, 0; x)
 \end{array} $
 &  $\begin{array}  {lcl}
P_{Alice}(x=b, b'=1)=\frac{1+f_1(\theta,~N=6)}{2}
 \end{array} $\\\hline
 1&  $\begin{array}  {lcl}
 {\mathcal P}^{A_1A_2}_{b'=1}(0, 0; x)
\end{array}$
& 
$\begin{array}  {lcl}
 {\mathcal P}^{B_1B_2}_{b'=1}(y, y; x)
 \end{array} $
 &
 $\begin{array}  {lcl}
 {\mathcal P}^{C_1C_2}_{b'=1}(0, c; x)
 \end{array} $
 &  $\begin{array}  {lcl}
P_{Bob}(y=c, b'=1)= \frac{1+f_1(\theta, ~N=6)}{2}
 \end{array} $\\\hline
 1&  $\begin{array}  {lcl}
 {\mathcal P}^{A_1A_2}_{b'=1}(0, a; x)
\end{array}$
& 
$\begin{array}  {lcl}
 {\mathcal P}^{B_1 B_2}_{b'=1}(0, 0; x)
 \end{array} $
 &
 $\begin{array}  {lcl}
 {\mathcal P}^{C_1 C_2}_{b'=1}(z, z; x)
 \end{array} $
 &  $\begin{array}  {lcl}
P_{Charlie}(z=a, b'=1)= \frac{1+f_1(\theta, ~N=6)}{2}
 \end{array} $\\\hline
 \end{tabular}
\caption{ List of quantum measurement operators ${\mathcal P}^{A_1A_2}_{b'}(\alpha, \beta;  \zeta)$, ${\mathcal P}^{B_1B_2}_{b'}(\alpha, \beta;  \zeta)$, ${\mathcal P}^{C_1C_2}_{b'}(\alpha, \beta;  \zeta)$ applied  on the process matrix $W$ expressed in Eq. (\ref{eqn:process_three_party}) for each strategy ($b'$)  which finally  yields $P^{Left}_{success}=\frac{1+f_0(\theta, ~N=6)}{2}$ and $P^{Right}_{success}=\frac{1+f_1(\theta, ~N=6)}{2}$. Hence, we get   $P_{success}=\frac{P^{Left}_{success}+P^{Right}_{success}}{2}=\frac{2+f_0(\theta, ~N=6)+f_1(\theta, ~N=6)}{4}$.}
\label{Table2}
\end{table}
\end{center}
\end{widetext}
By exploiting those strategies on the process matrix $W$ expressed in Eq. (\ref{eqn:process_three_party}), we find 
\begin{eqnarray}
P^{Left}_{success}&=&\frac{1+f_0(\theta, ~N=6)}{2},~ P^{Right}_{success}=\frac{1+f_1(\theta, ~N=6)}{2}. \nonumber\\
\label{en:success:new}
\end{eqnarray}
Using this  we finally get 
\begin{eqnarray}
P_{success}&=&\frac{P^{Left}_{success}+P^{Right}_{success}}{2}\nonumber\\&=&\frac{2+f_0(\theta, ~N=6)+f_1(\theta, ~N=6)}{4}.\nonumber\\
\label{eqn:success:three_party}
\end{eqnarray}
 However, we would like to mention that the process matrix we consider in Eq. (\ref{eqn:process_three_party}) is not the optimal one. It has been shown in Ref. \cite{causal_order_three_parties} that there exist a  process matrix and a set of measurements for which one can get the maximum value of the total success probability to be 1, which is higher than the value obtained for the two-player game.  Hence, one can think that in some sense the proposed three-party causal order game is analogous to the nonlocal game with the GHZ state \cite{GHZ1,GHZ2}. As similar to the maximal violation of causal order obtained in this case, in the GHZ game, the violation of locality is maximal. In our case, we do not consider such a process matrix as it is structurally very different from the quantum many-body Hamiltonian we consider in our work. The $W$ matrix considered in  Ref. \cite{causal_order_three_parties}  consists of four and five body terms while in our game to make it consistent with the quantum many-body Hamiltonian, the operators involved in $W$ are only two and three-body. 
 
Therefore, we argue that the original causal order game can be extended to a multi-player game consisting of $\mathcal{N}=N/2$ parties ($\mathcal{S}_1, \mathcal{S}_2,\dots, \mathcal{S}_\mathcal{N}$) such that if we arrange the parties in a clockwise cyclic pattern according to ascending order of their index, each of them has to guess the random classical bits of their immediate neighbors. We call $\mathcal{S}_i $ is the left neighbor of $\mathcal{S}_{i+1} $   and the right neighbor of $\mathcal{S}_{i-1}$, with boundary conditions $\mathcal{S}_0=\mathcal{S}_\mathcal{N}$   and  $\mathcal{S}_{\mathcal{N}+1}=\mathcal{S}_1$. If Crupier provides $b'=0$, everyone needs to guess the bit of their left neighbor ($P^{Left}_{\mathcal{S}_i}$) and $P^{Left}_{success}(b'=0)$ is just the average of the left probability computed for each party, i.e., $P^{Left}_{success}(b'=0)=\frac{1}{\mathcal{N}}\sum_{i=1}^{\mathcal{N}} P^{Left}_{\mathcal{S}_i}$. Similarly when $b'=1$,  the parties need to guess the random bit of their right neighbor ($P^{Right}_{\mathcal{S}_i}$) which will estimate  $P^{Right}_{success}(b'=1)=\frac{1}{\mathcal{N}}\sum_{i=1}^{\mathcal{N}}  P^{Right}_{\mathcal{S}_i}$. Now same as before,   we consider a definite causal ordering between the parties, such that $\mathcal{S}_1 \preceq \mathcal{S}_2 \preceq\mathcal{S}_3 \preceq \dots \preceq  \mathcal{S}_\mathcal{N}$. As a result, we find that for   $(\mathcal{N}-2)$ parties ($\mathcal{S}_2$ to $\mathcal{S}_{\mathcal{N}-1}$),  the  neighbor in the left remains in the causal past and thus they can guess its bit perfectly. Whereas, the other neighbor  is in their causal future. Hence, they can only guess its classical bit randomly. Therefore, together they   contribute factors  $\frac{\mathcal{N}-2}{\mathcal{N}}$ and $\frac{\mathcal{N}-2}{2\mathcal{N}}$ to  $P^{Left}_{success}(b'=0)$ and $P^{Right}_{success}(b'=1)$, respectively. However, if we consider $\mathcal{S}_1$, both of its neighbors remain in its causal future. Hence, it can only guess their bits randomly that yields $P_{\mathcal{S}_1}^{Left}=P_{\mathcal{S}_1}^{Right}=\frac{1}{2}$. Similarly,    $\mathcal{S}_{\mathcal{N}}$ remains in the causal future of its both  neighbors and can guess their bits exactly, that yields $P_{\mathcal{S}_\mathcal{N}}^{Left}=P_{\mathcal{S}_\mathcal{N}}^{Right}=1$. Therefore, we finally get 
  \begin{eqnarray}
  P^{Left}_{success}(b'=0)&=& 1-\frac{1}{2\mathcal{N}},  ~\text{and} ~P^{Right}_{success}(b'=1)= \frac{1}{2}+\frac{1}{2\mathcal{N}}. \nonumber\\
  \end{eqnarray}
  Hence, the total success probability of the classical game is again 
  $P_{success}^{classical}=\frac{1}{2} \Big(P^{Left}_{success}(b'=0)+P^{Right}_{success}(b'=1)\Big)=\frac{3}{4}$.

 In the quantum version of the $\mathcal{N}$-party game, we want the success probabilities  to take the values  similar to those derived  in Eqs. (\ref{en:success:new})-(\ref{eqn:success:three_party}), given by 
 \begin{eqnarray}
P^{Left}_{success}&=&\frac{1+f_0(\theta, ~N)}{2},~ P^{Right}_{success}=\frac{1+f_1(\theta, ~N)}{2},  
\label{en:success:N_party}
\end{eqnarray}
 and thus yielding 
\begin{eqnarray}
P_{success}&=&\frac{P^{Left}_{success}+P^{Right}_{success}}{2}=\frac{2+f_0(\theta, ~N)+f_1(\theta, ~N)}{4}.\nonumber\\
\label{eqn:total_success:N_party}
\end{eqnarray}
We argue that the above probabilities  can be obtained by exploiting  appropriate measurement strategy on an $\mathcal{N}$-party process matrix with the  general form  given by 
\begin{eqnarray}
W(\theta)&=&\frac{1}{2^{\mathcal{N}}}\Big[\mathbb{I}+\frac{f_0(\theta, ~N)}{\mathcal{N}_0}\Big( \sum_{i=1}^ {\mathcal{N}_0}\sigma_{\alpha}^{\mathcal{S}_i^{O}}\sigma_{\beta}^{\mathcal{S}_{i+1}^I}\Big)\nonumber\\&+&\frac{f_1(\theta, ~N)}{\mathcal{N}_1}\Big(\sum_{i=1}^ {\mathcal{N}_1}\sigma_{\gamma}^{\mathcal{S}_i^I} \sigma_{\delta}^{\mathcal{S}_{i+1}^I}\sigma_{\eta}^{\mathcal{S}^O_{i+1}}\Big)\Big],\nonumber\\
\label{eqn:process_N_party}
\end{eqnarray}
where $\mathcal{N}_0$ ($\mathcal{N}_1$)  is the  number of two-body (three-body) terms in $W$, $\mathcal{S}_i^I$ ($\mathcal{S}_i^O$) is the input (output) quantum system of the $i^{th}$ party. Whereas,    $\alpha, \beta, \gamma, \delta, \eta \in \{x, y, z\}$ and their choices depend on certain factors, e.g., number of parties, validity of $W$ as a process matrix, etc. For example, for $N=6$ ($\mathcal{N}=3$), we choose $\alpha=\beta=\eta=z$, $\gamma=\delta=x$ and $\mathcal{N}_0=\mathcal{N}_1=3$.

To establish a correspondence of this generalized game with a suitable quantum many-body system, we consider a quantum many-body Hamiltonian similar to the previous one that includes an Ising interaction and a cluster Hamiltonian, expressed as 
\begin{eqnarray}
\mathcal{H}(\theta)=-\cos\theta \sum_{i=1}^N\sigma_z^i \sigma_z^{i+2}-\sin\theta \sum_{i=1}^N \sigma_z^i \sigma_x^{i+1} \sigma_z^{i+2}, 
\label{eqn:hamiltonian_new}
\end{eqnarray}
where  PBC are enforced. 
One can note that for $N>4$, the number of terms present in the Ising part and the cluster part is the same. Hence, we no longer need the  factor 2 in front of the Ising part. The Hamiltonian can be solved exactly \cite{ref_new1,ref_new2} by first mapping it to two transverse field Ising  models using the controlled-phase ($\mathcal{CZ}$)  operation on pair of sites, given by  
\begin{eqnarray}
\mathcal{U}=\Pi_i \mathcal{CZ}_{i,i+1},\text{with}\hspace{0.1cm} \mathcal{CZ}_{i,i+1}=|0\rangle \langle 0|^i \otimes \mathbb{I}^{i+1}+|1\rangle \langle 1|^i \otimes \sigma_z^{i+1}, \nonumber\\
\label{en:op}
\end{eqnarray}
which yields 
\begin{eqnarray}
\mathcal{U} \mathcal{H}\mathcal{U}^{\dagger}=\mathcal{H}_{\text{TFIM}}^{odd}+\mathcal{H}_{\text{TFIM}}^{even},
\end{eqnarray}
where $\mathcal{H}^{odd}_{\text{TFIM}}$ and $\mathcal{H}^{even}_{\text{TFIM}}$ are the nearest-neighbor transverse field Ising model (TFIM)   on odd and even sites respectively, defined as 
\begin{eqnarray}
\mathcal{H}^{odd}_{\text{TFIM}}&=&-\cos \theta\sum_{i=1}^ {N/2} \sigma_z^{2i-1} \sigma_z^{2i+1}-\sin \theta \sum_{i=1}^{N/2}\sigma_x^{2i-1},\nonumber\\
\mathcal{H}^{even}_{\text{TFIM}}&=&-\cos \theta \sum_{i=1}^ {N/2} \sigma_z^{2i} \sigma_z^{2(i+1)}-\sin \theta \sum_{i=1}^{N/2}\sigma_x^{2i}.
\label{eqn:Ising}
\end{eqnarray}
Each of the  above TFIMs can be solved exactly  by mapping it to a free-fermionic model using Jordan-Wigner transformation \cite{JW} (see Appendix \ref{AppendixA}). Therefore, the transformation not only helps us to solve the model Hamiltonian given in Eq. (\ref{eqn:hamiltonian_new}) explicitly, it also relates the model to one with well-understood quantum phases and order parameters. Now as these two TFIMs of equal length do not interact with each other, it is enough to diagonalize one of them and compute the relevant physical quantities of the actual model out of that. Hence, from now onwards, we only consider one of them with the general form
\begin{eqnarray}
 \mathcal{H}_{\text{TFIM}}&=&-\cos \theta \sum_{i=1}^ {N/2} \sigma_z^{i} \sigma_z^{i+1}-\sin \theta \sum_{i=1}^{N/2}\sigma_x^{i},
 \label{eqn:Ising_second}
 \end{eqnarray}
 and let us denote the GS of this model by $|\Phi\rangle_g$. 
Moreover, under the unitary operation defined in Eq. (\ref{en:op})  the projector  in Eq. (\ref{eqn:operators})
  changes as 
\begin{eqnarray}
\Pi^0 = \frac{1}{2}\big(\mathbb{I}^{i~i+1~i+2}+\sigma_z^i \sigma_x^{i+1} \sigma_z^{i+2} \big)& \rightarrow  & \mathcal{U} \Pi^0 \mathcal{U}^{\dagger}=\frac{1}{2}\big(\mathbb{I}^{i+1}+\sigma^{i+1}_x)
\nonumber,\\
 \Pi^1 =\frac{1}{2}\big(\mathbb{I}^{i~i+2}+\sigma_z^i \sigma_z^{i+2}\big)& \rightarrow  &\mathcal{U} \Pi^1 \mathcal{U}^{\dagger}=\Pi^1.
\end{eqnarray}
Hence, the exact solution of the TFIMs provides a close analytical form of all the above quantities for any arbitrary system size $N$.

\begin{figure}[t]
\includegraphics[width=7.cm,angle=-0]{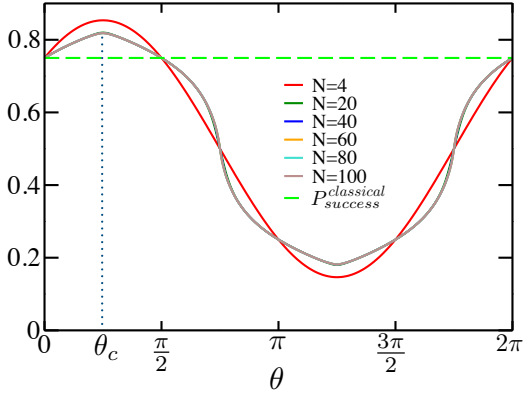}~~~
\caption{Plot of   scaling  of success probability of the $\mathcal{N}$-party causal order game with the process matrix defined in Eq. (\ref{eqn:process_N_party}) or equivalently  $\mathcal{K}_{avg}$ computed for the GS of the model defined in Eq. (\ref{eqn:hamiltonian_new}) (for $N>4$ and Eq. (\ref{eqn:hamiltonian}) for $N=4$).  The plot for    system size  $N=4$ is denoted by broken  red curve. Whereas,  all the remaining  plots  (i.e., for $N= 20, 40, 80, 100$) are denoted by solid curves.  One can note that the plots for $N\geq 20$ become indistinguishable from each other. Moreover, similar to $N=4$ case,  all of them attain maxima at $\theta_c=\frac{\pi}{4}$. The  horizontal broken straight line has the same meaning as in Fig. \ref{fig:sum_observable}}.
\label{fig:probability_N_party}
\end{figure}

 We now consider that the expectation value of the two-body and the three-body operators are  equal to the functions $f_i(\theta,N)$ introduced in Eq. (\ref{eqn:process_N_party}), i.e., 
\begin{eqnarray} 
 f_0(\theta, ~N)&=&C_{zz}={_g}\langle \Phi|\sigma_z^i \sigma_z^{i+2}|\Phi\rangle_g, \nonumber\\
 f_1(\theta, ~N)&=&m_{x}={_g}\langle \Phi| \sigma_x^i|\Phi\rangle_g.
\end{eqnarray}
 This again provides us with the scope to relate the success probabilities obtained for the $\mathcal{N}$-party causal order game with that of the expectation values of the projectors computed for the GS of the quantum many-body Hamiltonian of the size $N=2\mathcal{N}$.  In other words, we get  from Eqs. (\ref{en:success:new})-(\ref{eqn:success:three_party})
{\small 
\begin{eqnarray}
P_{success}^{Left}=\frac{1+C_{zz}}{2}, P_{success}^{Right}=\frac{1+m_x}{2}.\nonumber\\
\label{eqn:probability_observable}
\end{eqnarray}
}
Hence, the total success probability of the game can again be related with the average of the expectation values of the projectors ($\mathcal{K}_{avg}$), as introduced in Eq. (\ref{eqn:op_sum}), 
\begin{eqnarray}
\mathcal{K}_{avg}\equiv
P_{success}=\frac{2+m_x+C_{zz}}{4}. \nonumber\\
\end{eqnarray}
We plot the behavior of $\mathcal{K}_{avg}$ ($P_{success}$) in Fig. \ref{fig:probability_N_party} for different system sizes. One can note that similar to $N=4$ case, $\mathcal{K}_{avg}$ ($P_{success}$) becomes maximum at $\theta_c=\frac{\pi}{4}$ for all values of $N$. Hence, the correspondence observed for $N=4$ equally holds for large system size: maximum violation of the causality corresponds to a second-order QPT in the considered  quantum many-body model. 

\section{Detection OF  quantum phase transition point}
\label{sec5}

In this section, we provide a detailed discussion about efficient detection of the QPT point in the model Hamiltonian defined in Eq. (\ref{eqn:hamiltonian_new}). For that purpose, we first seek suitable order parameters characterizing the Ising and the cluster phases of the model and analyze their behavior with the parameter $\theta$.  Note that for our analysis, it is enough to choose the region of interest as  $0\leq \theta \leq \frac{\pi}{2}$. From Eq. (\ref{eqn:Ising_second}), we can see for the Ising phase, a suitable order parameter is the longitudinal magnetic field $m_z={_g}\langle \Phi| \sigma^i_z|\Phi\rangle_g$ which remains non-zero only for the region $0\leq \theta < \frac{\pi}{4}$.   In the thermodynamic limit, $m_z$ behaves as \cite{pfeuty}

\begin{figure}[t]
\begin{center}
\includegraphics[width=7.cm,angle=-0]{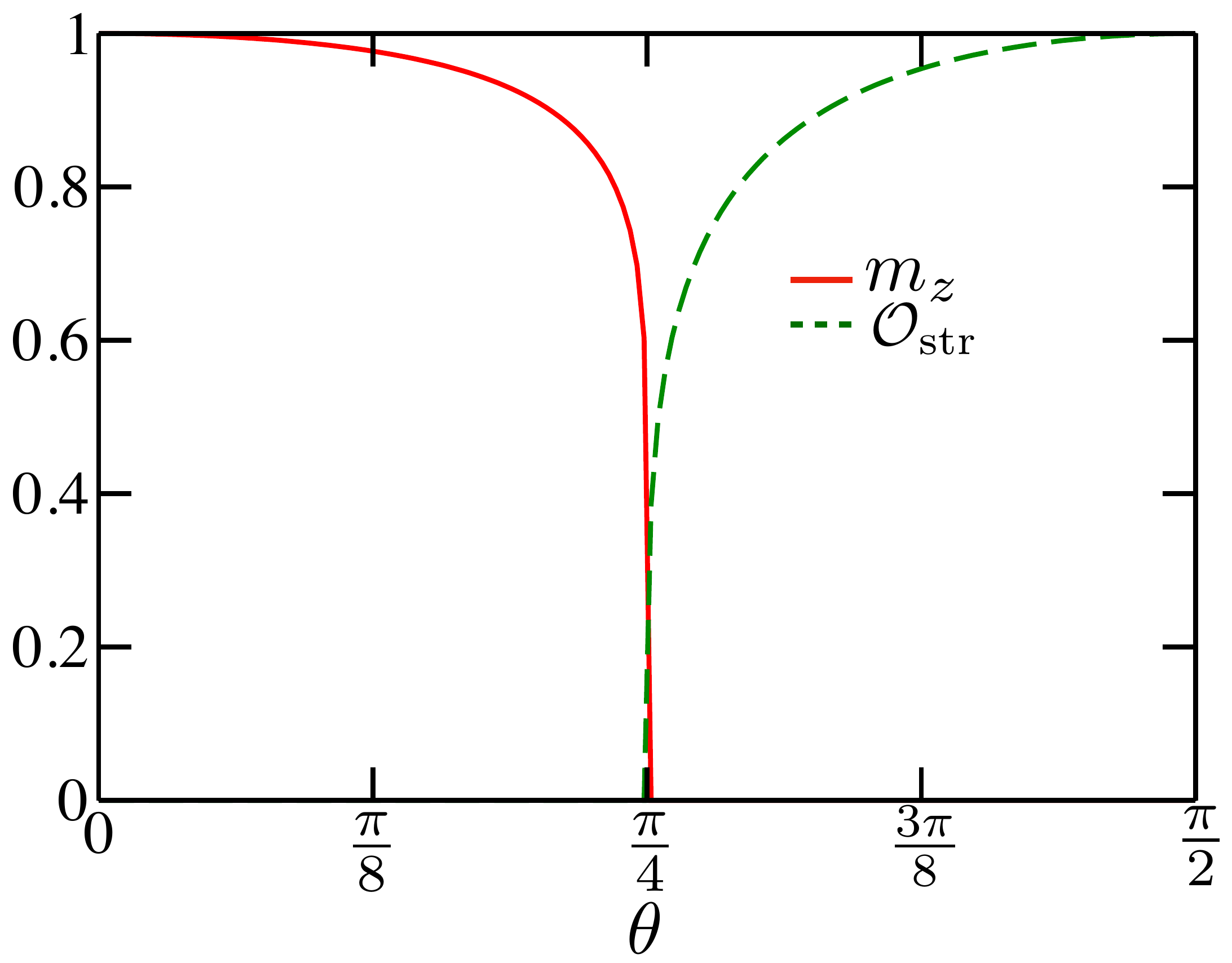}~~~
\end{center}
\caption{ Plot of   the behavior of the order parameters characterizing the Ising phase, $m_z$ (solid red line)  and the cluster phase,  $\mathcal{O}_{\text{str}}$ (broken green line) defined in Eq. (\ref{eqn:SOP})  as a function of $\theta$. $m_z$ remains non-zero for the region  $0\leq \theta < \frac{\pi}{4}$  and decays following  Eq. (\ref{eqn:mz_decay}). Whereas, $\mathcal{O}_{\text{str}}$  exhbits completely opposite behavior and has non-zero value only for the region $\frac{\pi}{4} < \theta \leq \frac{\pi}{2}$ where it decays following Eq. (\ref{eqn:SOP_decay}).}
\label{fig:SOP_mz}
\end{figure}

\begin{center}
\begin{equation}
m_z = 
\begin{cases}
  (1-\tan^2\theta)^{1/8} & \text{for  $0 \leq  \theta<\frac{\pi}{4}$}, \\
  0 & \text{for  $\frac{\pi}{4}  <  \theta \leq\frac{\pi}{2}$.}
\end{cases}
\label{eqn:mz_decay}
\end{equation}
\end{center}
On the other hand, for the cluster phase we choose the following string order parameter 
that characterizes hidden antiferromagnetic order of the model, defined as \cite{SOP1, pollmann}
\begin{eqnarray}
\mathcal{O}_{\text{str}}&=&{_g}\langle \Psi| \Pi_{i=1}^N  \sigma_x^i| \Psi \rangle_g. 
\label{eqn:SOP}
\end{eqnarray}
Now it is easy to see that $\Pi_{i=1}^{N} \sigma_x^i$ remains invariant under the transformation given in Eq. (\ref{en:op}), i.e., 
\begin{eqnarray}
\mathcal{U} \Pi_{i=1}^{N} \sigma_x^i \mathcal{U}^{\dagger}= \Pi_{i=1}^N \sigma_x^i. 
\end{eqnarray}
Hence, to compute the value of $\mathcal{O}_{\text{str}}$ for the GS of the TFIM we need to consider the same string of operators,
\begin{eqnarray}
\mathcal{O}_{\text{str}}&=& \Big[{_g}\langle \Phi| \Pi_{i=1}^{N/2} \sigma_x^i|\Phi\rangle_g\Big]^2.
\end{eqnarray}
Moreover,  using  the Kramers-Wannier duality transformation \cite{Kramers}, 
\begin{eqnarray}
\mu^{i}_z=\Pi_{j=1}^i \sigma^j_x, \hspace{.2cm}\mu^i_x=\sigma^i_z \sigma^{i+1}_z, 
\end{eqnarray}
one can define a dual Hamiltonian to the TFIM model expressed in Eq. (\ref{eqn:Ising_second}) and get
\begin{eqnarray}
H_{\text{TFIM}}^{dual}=-\sin \theta \sum_{i=1}^{N/2} \mu_z^i \mu^{i+1}_z-\cos \theta \sum_{i=1}^{N/2} \mu_x^i.
\end{eqnarray}
This suggests,  computation of the string operator $\Pi_{i=1}^{N/2}\sigma_x^i$ for the GS of the TFIM defined in Eq. (\ref{eqn:Ising_second}) is equivalent to computation of the longitudinal magnetization ($\langle \mu^{N/2}_z\rangle$) for the GS of the dual Hamiltonian $H_{\text{TFIM}}^{dual}$. In other words,
\begin{eqnarray}
\mathcal{O}_{\text{str}}&=& \Big[{_g}\langle \Phi^{dual}|\mu_z^{N/2}|\Phi^{dual}\rangle_g\Big]^2, 
\end{eqnarray}
where we denote  $|\Phi^{dual}\rangle_g$ as the GS of $H_{\text{TFIM}}^{dual}$. Again, using the analytical form of the longitudinal magnetization in the thermodynamic limit, we get 
\begin{center}
\begin{equation}
\mathcal{O}_{\text{str}} = 
\begin{cases}
 0 & \text{for  $0 \leq  \theta<\frac{\pi}{4}$}, \\
   (1-\cot^2\theta)^{1/4} & \text{for  $\frac{\pi}{4}<   \theta  \leq\frac{\pi}{2}$.}
\end{cases}
\label{eqn:SOP_decay}
\end{equation}
\end{center}

Therefore, from Eqs. (\ref{eqn:mz_decay}) and Eqs. (\ref{eqn:SOP_decay}) one can find that for the considered range of $\theta$ the behavior of these two quantities remain complementary to each other and thus they serve as efficient order parameters characterizing two different phases of the model. We provide a pictorial depiction of the same in Fig. \ref{fig:SOP_mz} where the QPT  point $\theta_c=\frac{\pi}{4}$ is marked by the point where both the quantities vanish simultaneously.

\begin{figure}[t]
\includegraphics[width=7.cm,angle=-0]{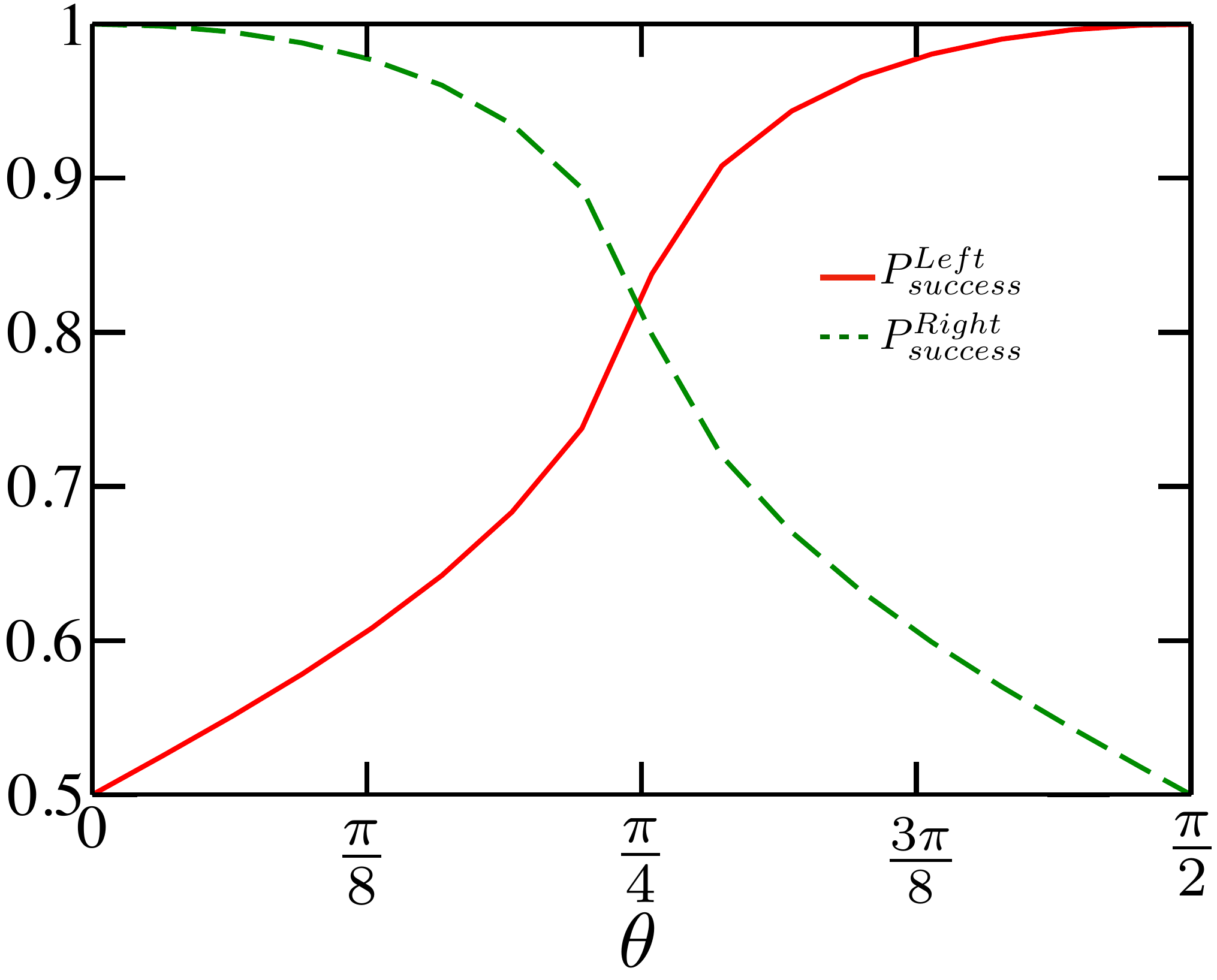}~~~
\caption{Plot of  the probabilities of the causal order game $P_{success}^{Left}$  (broken green line) and $P_{success}^{Right}$  (solid red line) which are the functions of  the observables $C_{zz}$ and $m_x$ respectively (see Eq. (\ref{eqn:probability_observable})),  computed for the GS of the TFIM defined in Eq. (\ref{eqn:Ising_second}) with $\theta$. Here we consider $N=100$.}
\label{fig:probability_theta}
\end{figure}

We now compare the success probabilities obtained for an $N$-party causal order game with the different phases of the considered Hamiltonian for the region $0\leq \theta \leq \frac{\pi}{2}$. For that purpose, we plot the behavior of the success probabilities ($P_{success}^{Left/Right}$) obtained using Eq. (\ref{eqn:probability_observable}) for large value of $N$ in Fig. \ref{fig:probability_theta} and compare that with Fig. \ref{fig:SOP_mz}. From the plot, we note that different phases of the model favour different strategies of the game. For instance, for the region $0 \leq \theta < \frac{\pi}{4}$ when the system remains  in  the Ising phase, $P_{success}^{Left}$ dominates over   $P_{success}^{Right}$. Similarly, for $\frac{\pi}{4} <\theta \leq \frac{\pi}{2}$ when the system remains in the SPT phase,  $P_{success}^{Right}$  becomes higher than $P_{success}^{Left}$. At the QPT point,  we get $P_{success}^{Left}=P_{success}^{Right}$ and  the total success probability becomes maximum. 

  \section{Discussion and future work}
  \label{sec6}
  In this work, we have presented a framework to relate the eigenstates of a topological Hamiltonian with the resource of causal order game introduced in Ref. \cite{brukner}, called the process matrix. In particular, we have shown that the success probabilities of different strategies of the game can be realized as expectation values of different observables computed for the eigenstates of the model.  Thus the GS and the most excited state can be related to a non-causally separable process matrix whenever the sum of a certain two-body and three-body correlations exceed a minimum value. However, for other excited states, the process matrix to which they can be related remains causally separable for the whole range of the system parameters. In addition to this, we showed that at the point of maximum violation of the causal order, in the thermodynamic limit  corresponding quantum many-body model undergoes a second-order QPT.   We further showed that different quantum phases of the model can favor different quantum strategies of the model. For instance, in the Ising phase $P_{success}^{Left}$ remains higher than that of $P_{success}^{Right}$. Whereas, for the cluster phase of the model opposite ordering appeared.  The results have been generalized for a higher number of parties and the behavior remains the same. 
  We believe that our work is a  genuine attempt to establish a  correspondence between quantum many-body systems and quantum game theory which may become useful for experimental realization of the game-theoretic schemes \cite{IBM}. As future work, we wish to consider a generalized version of the Hamiltonian  $\mathcal{H}$ that includes additional non-commutative terms along with long-range interactions and relate that to a modified version of the game.

\acknowledgments
A. Bera thanks G. Sierra  and  IFT, Madrid, Spain  for the visit and hospitality during which most of the work of this project have been  carried out.  G. Sierra  thanks Titus Neupert for the invitation to the Mini workshop on Quantum Computing, 
Zurich, May 2019. We also acknowledge conversations with Nicolas Regnault, Frank Pollmann, David P{\'e}rez-Garc{\'i}a,   Antonio Ac{\'i}n,  Esperanza L{\'o}pez and Maciej Lewenstein. We thank  Javier Rodr{\'i}guez-Laguna for reading the manuscript and providing useful suggestions.  We also thank the anonymous referees for providing useful suggestions that have helped us to improve the manuscript. The article has received  financial support from the
grants PGC2018-095862-B-C21,  QUITEMAD+ S2013/ICE-2801, SEV-2016-0597 of the
{\em Centro de Excelencia Severo Ochoa} Programme,  the CSIC
Research Platform on Quantum Technologies PTI-001 and the Polish National Science Centre project 2018/30/A/ST2/00837.

\appendix

\begin{widetext}
\section{Jordan-Wigner transformation and computation of observables}
\label{AppendixA}
We start with the nearest-neighbor transverse filed Ising model
\begin{eqnarray}
 \mathcal{H}_{\text{TFIM}}&=&-\cos \theta \sum_{i=1}^ {N/2} \sigma_z^{i} \sigma_z^{i+1}-\sin \theta \sum_{i=1}^{N/2}\sigma_x^{i}. 
 \label{eqn:Ising_third}
 \end{eqnarray}
 
 Now let us do the following transformation 
 \begin{eqnarray}
 \sigma_x^i=2c_i^{\dagger} c_i-1, \sigma_z^i=-\Pi_{k<i} (1-2c_k^{\dagger} c_k) (c_i+c_i^{\dagger}),
 \end{eqnarray}
 where $c_i (c_i^{\dagger})$ are spinless fermionic anhilation  (creation) operator at site $i$. 
 Plugging this in Eq. (\ref{eqn:Ising_third}) we get
 \begin{eqnarray}
 \mathcal{H}_{\text{TFIM}}&=&-\cos \theta \sum_{i=1}^ {N/2} \Big[c_i^{\dagger} c_{i+1}+c_{i+1}^{\dagger} c_{i}+c_{i}^\dagger c_{i+1}^\dagger +c_{i+1} c_{i}\Big]-\sin \theta \sum_{i=1}^{N/2}(2 c_i^{\dagger} c_{i}-1), \\
&=&\sum_{i,j =1}^ {N/2} \Big[c_i^{\dagger} A_{ij} c_{j}+\frac{1}{2}(c_{i}^{\dagger} B_{ij} c_{j}^{\dagger}+c_{i} B_{ij} c_{j})\Big],
 \end{eqnarray}
 where $A_{ij}=-\cos\theta (\delta_{j,i+1}+\delta_{i,j+1})-2\sin\theta \delta_{i,j}, B_{ij}=-\cos\theta(\delta_{j,i+1}-\delta_{i,j+1})$, with $A_{1N}=A_{N1}=\cos\theta$ and $B_{1N}=-B_{N1}=-\cos\theta$. 
 Now defining following set of vectors $\psi_k$ that obey the  eigenvalues equations 
 \begin{eqnarray}
 (A+B) (A-B)\psi_k^T=\Lambda_k^2 \psi_k^T,
 \end{eqnarray}
 and obtaining the corresponding $\phi_k$ using 
 \begin{eqnarray}
 \phi_k^T=\Lambda_k^{-1}(A-B)\psi_k^T,
 \end{eqnarray}
 we can define correlation matrix $G_{ij}$ as follows
 \begin{eqnarray}
 G_{ij}=-\sum_{k} \psi_{ki} \phi_{kj}=(\psi_k^T \phi_k)_{ij}.
 \end{eqnarray}
 This finally yields
 \begin{eqnarray}
 m^i_x=-G_{i,i}, C^{i,i+1}_{zz}=G_{i,i+1}.
 \label{eqn:mag_corr}
 \end{eqnarray}
 One can note that due to translational invariance the RHS of the above equations does not depend on the site index and we can write  $m_x=m_x^i$ and $C_{zz}=C_{zz}^{i,i+1}$. 

\end{widetext}


\begin{thebibliography}{99}
\bibitem{process_mat_def}
A process matrix $W$ is a positive semidefinite operator that acts on the tensor product of the input and output Hilbert spaces of Alice  and Bob. The set of conditions required for a matrix to be a valid process matrix  for the cases $W \in \mathcal{H}_{A_1} \otimes \mathcal{H}_{A_2}  \otimes \mathcal{H}_{B_1}  \otimes \mathcal{H}_{B_2}$ are given by




\begin{eqnarray}
W&\geq &0, \nonumber\\
\text{Tr}(W)&=& D, \nonumber\\
_{B_1B_2}W &=&_{A_2B_1B_2}W,  \nonumber\\
_{A_1A_2}W &=&_{A_1 A_2 B_2}W,  \nonumber\\
W &=& _{B_2}W + _{A_2}W-_{A_2B_2}W. 
\end{eqnarray}
where $D$ is the dimension of the output system, i.e., $D=$dim($A_2B_2$) and  the operator $\mathcal{X}$ denotes the CPTP map that can be realized as tracing out the subsystem $\mathcal{X}$ and replacing it by the normalized identity operator, defined as
$_{\mathcal{X}}W=\frac{\mathbb{I}^{\mathcal{X}}}{\mathcal{D}} \otimes  \text{Tr}_{\mathcal{X}} W$, with $\mathcal{D}=\text{dim}(\mathcal{X})$. The notion of process matrix is a generalization of the concept of quantum state. When we discard the output systems ($A_2$ and $B_2$), $W$ reduces to a valid density matrix characterizing a quantum system.
\end{thebibliography}
\end{document}